\documentclass[preprint2]{aastex}

\slugcomment{To appear in ApJ}

\shorttitle{Chemistry of the Trailing Arm of Sgr}
\shortauthors{Keller, Yong, \& Da Costa}

\begin{document}

\title{The Chemistry of the Trailing arm of the Sagittarius Dwarf Galaxy}


\author{Stefan C. Keller, David Yong and Gary S. Da Costa}
\affil{Research School of Astronomy and Astrophysics, Australian National University, \\Mt. Stromlo Observatory, Cotter Rd. Weston ACT 2611 Australia.}
\email{stefan@mso.anu.edu.au}


\begin{abstract}
We present abundances of C, O, Ti, and Fe for eleven M-giant stars in the trailing tidal arm of the Sagittarius dwarf (Sgr). The abundances were derived by comparing synthetic spectra with high-resolution infrared spectra obtained with the Phoenix spectrograph on the Gemini South telescope. The targeted stars are drawn from two regions of the Sgr trailing arm separated by 66$^{\circ}$ (5 stars) and 132$^{\circ}$ (6 stars) from the main body of Sgr. The trailing arm provides a more direct diagnostic of the chemical evolution of Sgr compared to the extensively phase-mixed leading arm. 

Within our restricted sample of $\sim$2--3 Gyr old stars, we find that the stream material exhibits a significant metallicity gradient of $-(2.4\pm0.3)\times10^{-3}$ dex / degree ($-(9.4\pm1.1)\times10^{-4}$ dex / kpc) away from the main body of Sgr. The tidal disruption of Sgr is a relatively recently event. We therefore interpret the presence of a metallicity gradient in the debris as indicative of a similar gradient in the progenitor. The fact that such a metallicity gradient survived for almost a Hubble time indicates that the efficiency of radial mixing was very low in the Sgr progenitor.

No significant gradient is seen to exist in the [$\alpha$/Fe] abundance ratio along the trailing arm. Our results may be accounted for by a radial decrease in star formation efficiency and/or radial increase in the efficiency of galactic wind-driven metal loss in the chemical evolution of the Sgr progenitor. The [Ti/Fe] and [O/Fe] abundance ratios observed within the stream are distinct from those of the Galactic halo. We conclude that the fraction of the intermediate to metal-rich halo population contributed by the recent dissolution ($<3$ Gyr) of Sgr-like dwarf galaxies can not be substantial. 
\end{abstract}





\section{Introduction}

The paradigm of $\Lambda$CDM cosmology enshrines the importance of hierarchical assembly to the process of galaxy formation. Galaxies such as the Milky Way are expected to arise from the coalescence of numerous smaller systems. The idea that the stellar halo of the Milky Way was formed from the disruption of smaller systems was first proposed by \citet{SZ}. The study of \citet{Carollo07} argues that the relative contribution of stars originating in external systems to the halo is dependent on Galactocentric radius ($R_{GC}$). The outer halo ($R_{GC}$ $>15$ kpc) is dominated by lower metallicity material and exhibits kinematics indicative of derivation from accretion. On the other hand, at $R_{GC}$ $<10$ kpc the inner halo is dominated by generally higher metallicity and prograde kinematics as would arise from an \citet{ELS} in-situ formation scenario. Further, observational studies of the spatial \citep{Bell08} and kinematic \citep{Starkenburg09} properties of the halo conclude that accretion has a significant role to play in the construction of the Halo. These studies find accretion is responsible for between 10-100\% of the extant population depending on the assumed merger tree for the Galaxy.

An apparent argument against substantial contribution of accreted stars to the Halo is the discrepancy between the chemistry of the most-likely accreted systems, namely,  dwarf spheroidal (dSphs) galaxies, and the chemistry of the Halo \citep[see for example,][and recently, \citeauthor{Tolstoy09} \citeyear{Tolstoy09}]{Unavane96}. At [Fe/H]$ > -1$ dex, dSph stars exhibit substantially lower [$\alpha$/Fe] ratios compared to their Galactic halo counterparts \citep{Tolstoy03,Shetrone03,Venn04}. Chemical evolution models \citep{Gilmore91, Lanfranchi04, Matteucci08, Calura09} demonstrate this is due to substantially different star formation histories, combined with differences in the ability to retain processed gas, between the Milky Way and dSph populations. A solution to the chemical mismatch is that the stellar populations of the present-day dSphs are not representative of the bulk of the material previously contributed to the halo via tidal stripping. Indeed the different chemistry between the present-day dSphs and the halo might simply reflect that these dSphs were able to survive for a Hubble time and this has provided an extended timescale for chemical evolution to occur within them \citep{Lagos09}.

As pointed out by \citet{Abadi06} and \citeauthor{Font06} (\citeyear{Keller06}), tidal disruption of a dSph in which there is a strong metallicity gradient would lead to the deposition of stars distinct from those surviving in the present-day core. In low mass systems, where orbit swapping processes are expected to be much less effective (\citeauthor{Roskara08} \citeyear{Roskara08}), and distinct from low mass early-type galaxies that have suffered major mergers (\citeauthor{Spolaor09} \citeyear{Spolaor09}), the sense of this metallicity gradient is one of decreasing metallicity with increasing radial distance \citep{Stinson09}. Furthermore, it has become apparent that the chemistry of some of the most metal-poor ([Fe/H]$ < -3$ dex) dSph stars is indistinguishable from that of the Halo \citep{Frebel10a, Norris10}. Hence tidal stripping of the metal-poor outskirts of a dSph could lead to the deposition of material that is old, metal-poor and exhibits abundance ratios matching those of the Halo \citep{Cohen09,Munoz05, Majewski03}.

In this paper we examine the chemistry of the stellar population that has been deposited into the outer halo of the Galaxy from the Sagittarius dwarf galaxy \citep[Sgr,][]{Ibata94}. The tidal disruption of the Sgr dwarf galaxy is the Milky Way's most prominent ongoing accretion event. Tidal debris on leading and trailing arms are traced via RR Lyrae variables \citep{Vivas04, Keller08, Prior09}, blue horizontal branch stars \citep{Yanny00,Yanny09,Newberg07}, M-giants \citep{Yanny09, Majewski03}, subgiants \citep{Keller09} and main-sequence turn-off stars \citep{fieldofstreams, Juric08}. The tidal debris extends around the sky and represents multiple orbits of Sgr around the Milky Way. 

This study utilises high resolution IR spectroscopy obtained on the Gemini South telescope to target stars in two regions along the length of the Sgr trailing arm. This enables us to determine the chemistry of material lost on two consecutive perigalactic passages of Sgr and hence to investigate the presence of any chemical abundance gradients evident in the stripped material.

\section{Sample Selection}

\begin{figure}
\begin{center}
\epsscale{0.80}
\includegraphics[width=95mm]{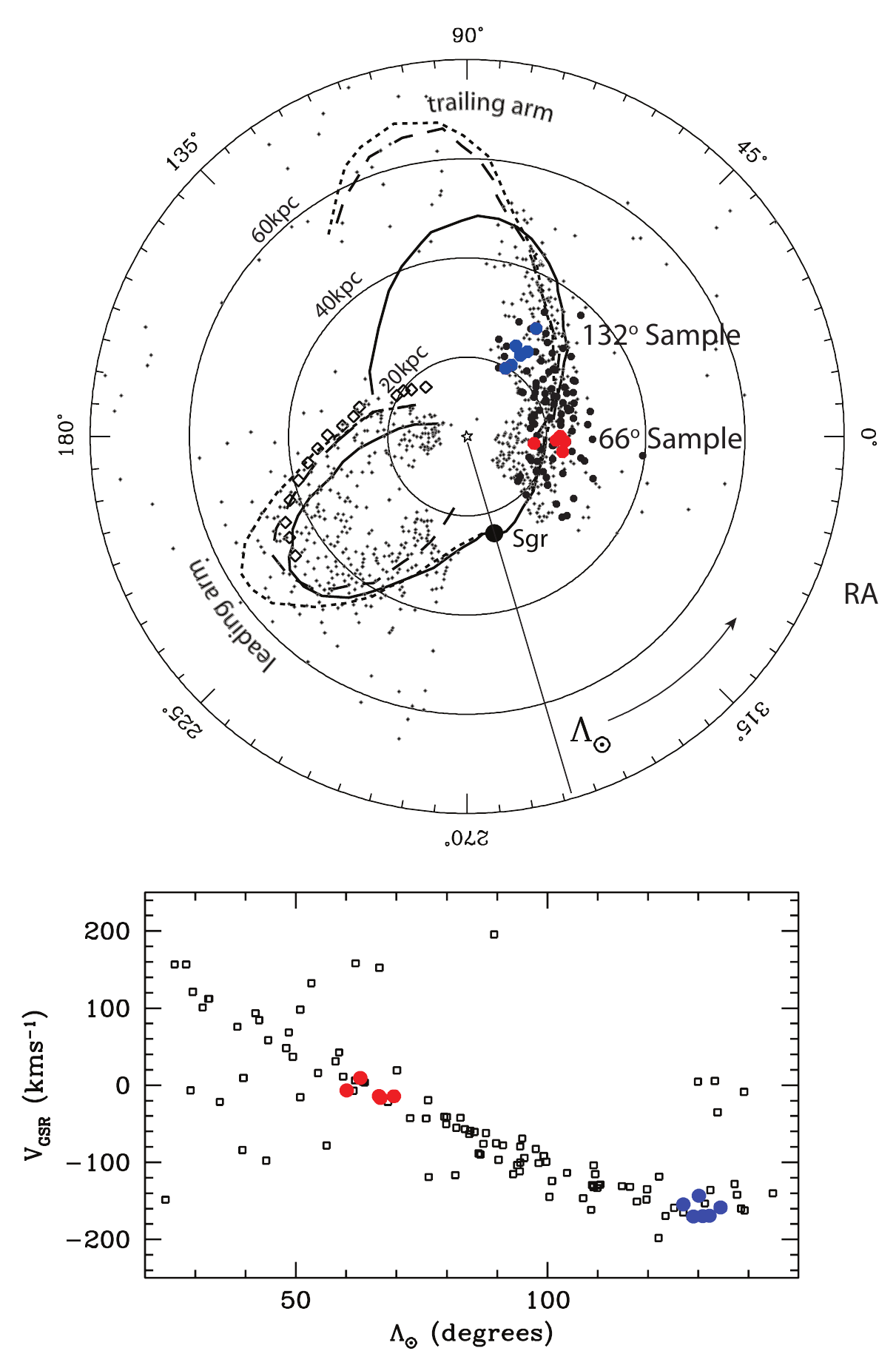}
\caption{{\bf{Top}}: The spatial distribution of target stars in the debris stream of Sgr ({\it{red}}: $\Lambda=66^{\circ}$ and {\it{blue}}:  $\Lambda=132^{\circ}$) projected in RA. Concentric circles show heliocentric distances of 20-80 kpc. M-giants from \citet{Majewski03} ({\it{bold black}}: radial velocities present and appropriate for Sgr material) and detections from \citet{fieldofstreams} ({\it{boxes}})) are shown. Overlaid are models from Law et al.
(2005, solid line), Fellhauer et al. (2006, dashed), and Helmi \& White (2001, short dashed).
{\bf{Bottom}}: The galactocentric radial velocities of the selected targets. Radial velocities are as determined by \citet{Majewski03}.
}\label{fig:spatial}
\end{center}
\end{figure}

The targets for our abundance study are taken from the M-giants catalogued in \citet{Majewski03}. We selected two samples of stars seen towards the trailing arm of Sgr debris that possess radial velocities and distances appropriate for the trailing arm. The radial velocities of the target sample are shown in Figure \ref{fig:spatial}. The two samples at $\Lambda_{\odot} = 66^{\circ}$ and $132^{\circ}$ are shown in Figure \ref{fig:spatial} in the Sgr plane in the context of literature models for the Sgr debris \citep{Law05,Fellhauer06, Helmi01}. As seen in the simulations of \citet{Law05} the two regions represent stars lost from the Sgr system approximately 0.5 Gyr and 1.3 Gyr ago respectively.

We have chosen to focus on the trailing arm for the following reasons. Firstly, \citet{Law05} show that the leading arm debris is much less spatially differentiated with respect to the epoch at which material was stripped compared with the trailing arm material (see their Figure 1). Consequently, stars lost in successive orbits overlap significantly. Furthermore, the radial velocity distribution of leading arm stars is less coherent than that seen in the trailing arm (see \citeauthor{Law05} \citeyear{Law05} Figure 10). For trailing arm material selection on the basis of radial velocity is more stringent and contamination from field star interlopers is less likely than that for a corresponding leading arm sample.

\section{Spectroscopic Observations}
Table \ref{table:obslog} reports the details of our observations. The spectra of individual stars along the trailing arm of Sgr were obtained in the $H$-band at a resolution of $R=50000$. The spectra were obtained using the 8 m Gemini South telescope and the Phoenix spectrograph \citep{Hinkle03} in queue mode observing (Program GS-2008B-Q-33). The observations were centred at 1.555 $\mu$m and utilised the H6420 order separating filter to provide a wavelength coverage of 75\AA. The spectra were observed using a 4-pixel slit (0.34$^{\prime\prime}$ wide and 14$^{\prime\prime}$ long). Each target star was observed at two positions, left and right of the midpoint of the spectrograph slit length, separated by 2.5$^{\prime\prime}$. The sky and dark background were removed by subtracting exposures taken at the alternating positions on the detector array. For each night 10 dark and flat frames were also acquired. A series of spectra of hot stars for telluric line correction were also acquired. Examination of the observed wavelength window reveals only a very few weak telluric features.

The frames were reduced in the IRAF environment following a procedure similar to that described in \citeauthor{Smith02} (\citeyear{Keller02}) and \citet{Melendez03}. Dark and flat frames are combined and the resulting dark is subtracted from the flat. A response image was derived from the flat and the science frames were then divided through by this response frame. The spectra were then extracted and wavelength calibrated using the stellar absorption lines evident. Finally, the spectra corresponding to left and right displacements were combined and continuum normalised.

\section{Stellar Parameters and Abundance Determination}
The reddening to each object is derived from the 100 $\mu$m Galactic map of \citet{Schlegel98} and is provided in Table \ref{table:obslog}. Infra-red photometry is derived from the 2MASS database having first transformed from the 2MASS system to the Johnson-Glass system\footnote{http://www.astro.caltech.edu/$\sim$jmc/2mass/v3/transformations/} for the purpose of temperature determination. We dereddened the colours of the target stars using a ratio of $E(J-K)/E(B-V) = 0.53$ \citep{Bessell98}. The effective temperature scale for cool giants is well defined from occulation measurements. We use the relation of \citet{Bessell83}, namely T$_{eff}$ = $7070 / (J-K + 0.88)$, to derive the temperatures given in Table \ref{table:starpars}. Surface gravities were derived from interpolation of appropriate low metallicity isochrones \citep{PadovaIso}. The microturbulent velocity was determined using the following relation, $\xi_{t} = 4.2-(6\times10^{-4}T_{eff})$, adopted from the optical analysis by \citet{Melendez08} of thick disk and bulge stars. We estimate that internal uncertainties in the stellar parameters are $T_{eff} \pm 75 K$, logg $g$ $\pm$ 0.4 dex, and $\xi_{t} \pm 0.4$ km s$^{-1}$.

\begin{figure*}
\includegraphics[width=140mm]{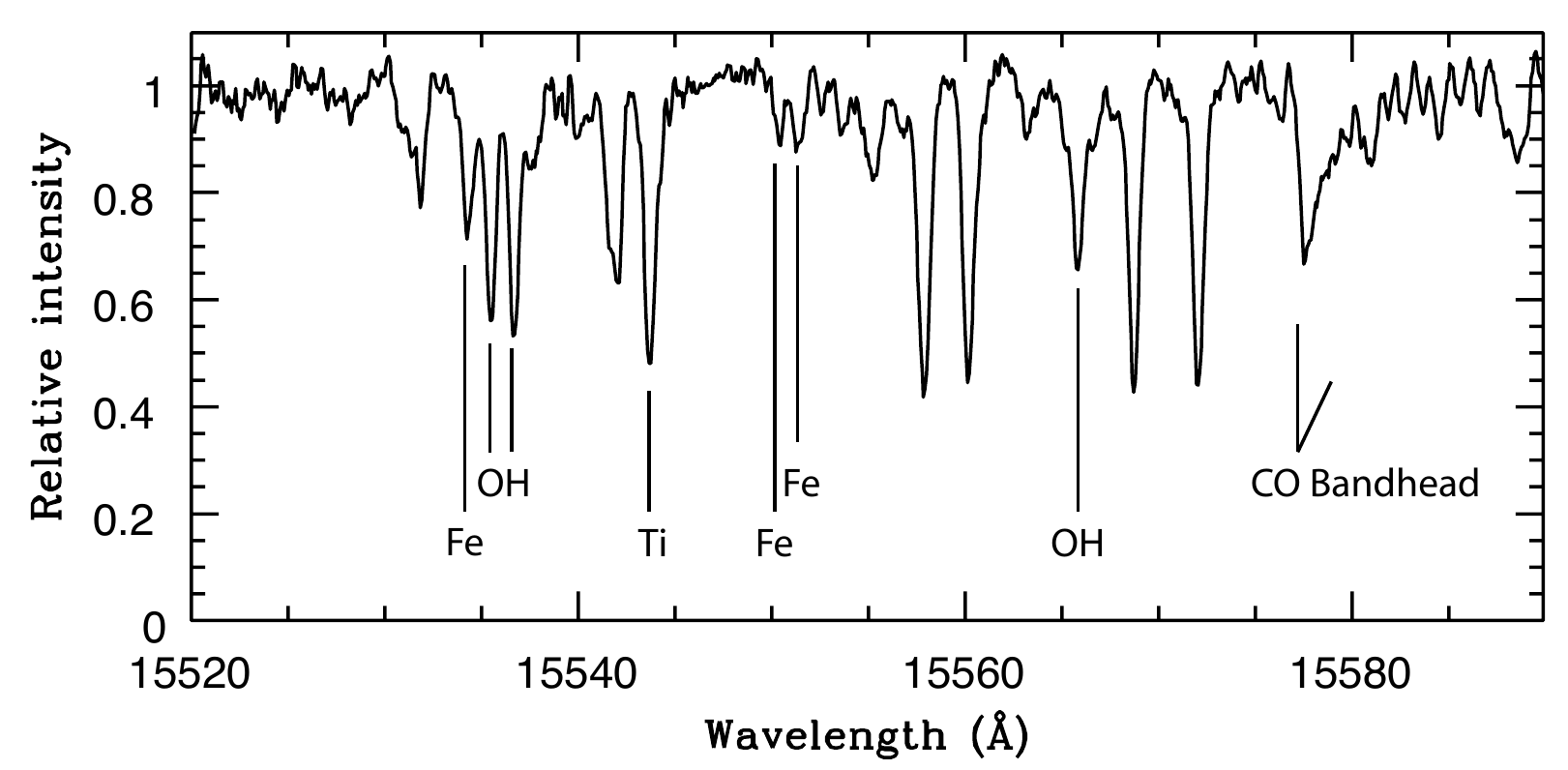}
\caption{Observed spectrum of 2350361m200216. Lines used in the abundance analysis are indicated.}\label{fig:spectrum_all}
\end{figure*}

\begin{figure}
\begin{center}
\includegraphics[width=84mm]{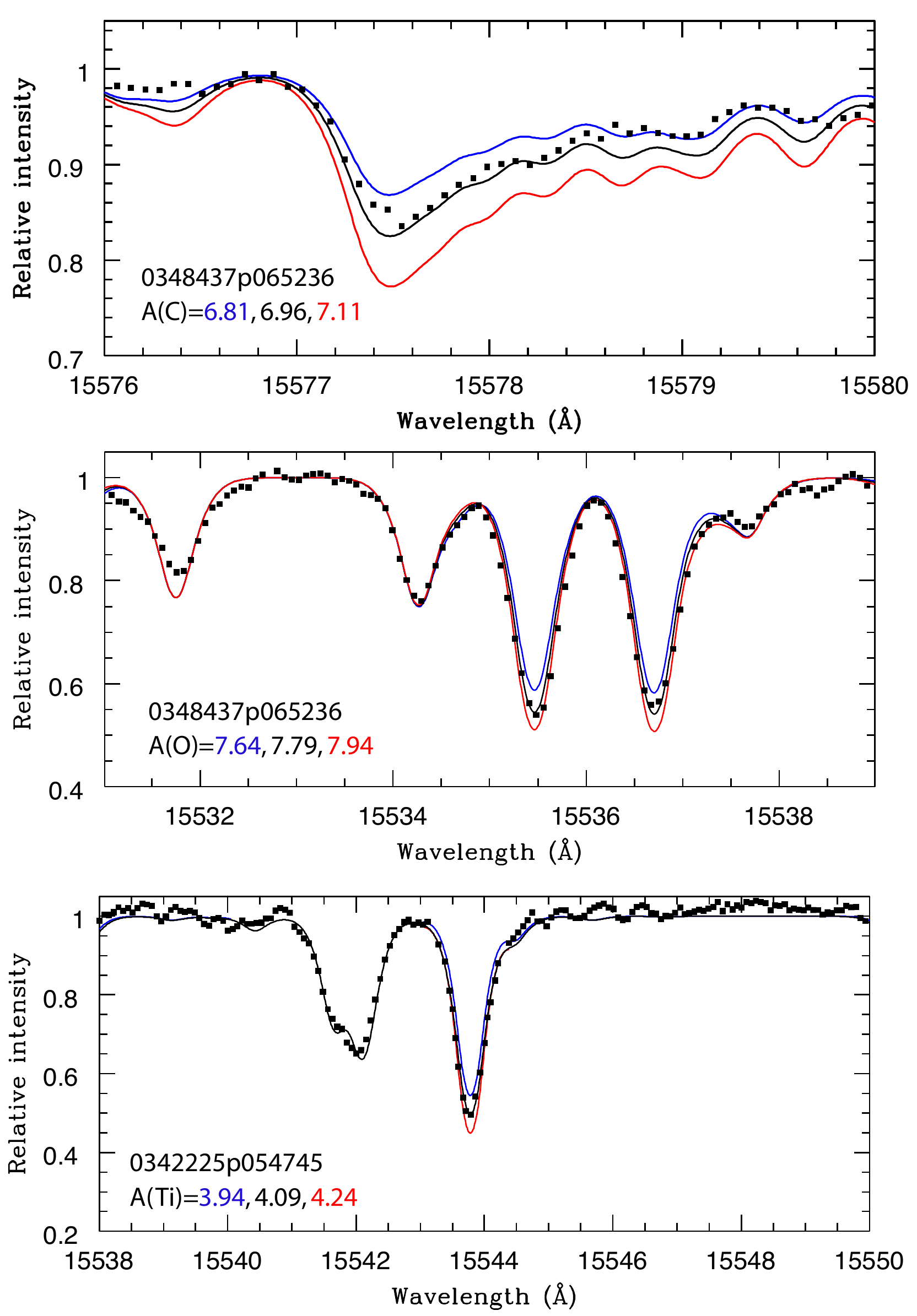}
\caption{Observed spectra ({\it{squares}}) and synthetic spectra for C ({\it{top}}); O ({\it{middle}}); and Ti ({\it{bottom}}). The synthetic spectra show the best fit ({\it{black line}}) and unsatisfactory fits ({\it{red and blue lines}}), $A$(C,O,Ti) $\pm$ 0.15 dex.}\label{fig:COTi}
\end{center}
\end{figure}

Abundances for a given line were derived by comparing synthetic spectra with the observed spectra following the analysis of \citet{Yong08}. The synthetic spectra were generated using the local thermodynamic equilibrium (LTE) stellar line analysis program MOOG \citep[][2007 version]{Sneden73} and LTE model atmospheres from the MARCS grid \citep{Gustafsson08}. The line list used in the generation of synthetic spectra was taken from \citet{Melendez99, Melendez01, Melendez03}. First we derived abundances for O from the OH molecular lines at 15535.462, 15536.705, and 15565.880 \AA (see Figure \ref{fig:spectrum_all}). Next, abundances for C were obtained from the CO molecular lines near 15576 \AA.  Since the abundances of C and O are coupled, we iterated until self-consistent abundances were obtained, which always occurred within one iteration. N measurements are possible from CN lines in the H-band. However, examination of the spectra revealed that reliable N abundances cannot be obtained given the weakness of the CN lines and the modest S/N ratios. Fe abundances were obtained from the Fe I lines at 15534.260\AA, 15550.450\AA, and 15551.430\AA. Ti abundances were determined from the Ti I line at  15543.780 \AA. In Figures \ref{fig:COTi}, we show examples of synthetic spectra fits to derive abundances in our sample, and in Table \ref{table:abundances} we present the final abundances. In the discussion to follow we express our abundances relative to the solar values of \citet{Asplund06}. The abundance dependences on the stellar parameters are given in Table \ref{table:abundanceUncertainties}.

\section{Metallicity Gradient} 
\begin{figure}
\begin{center}
\includegraphics[width=84mm]{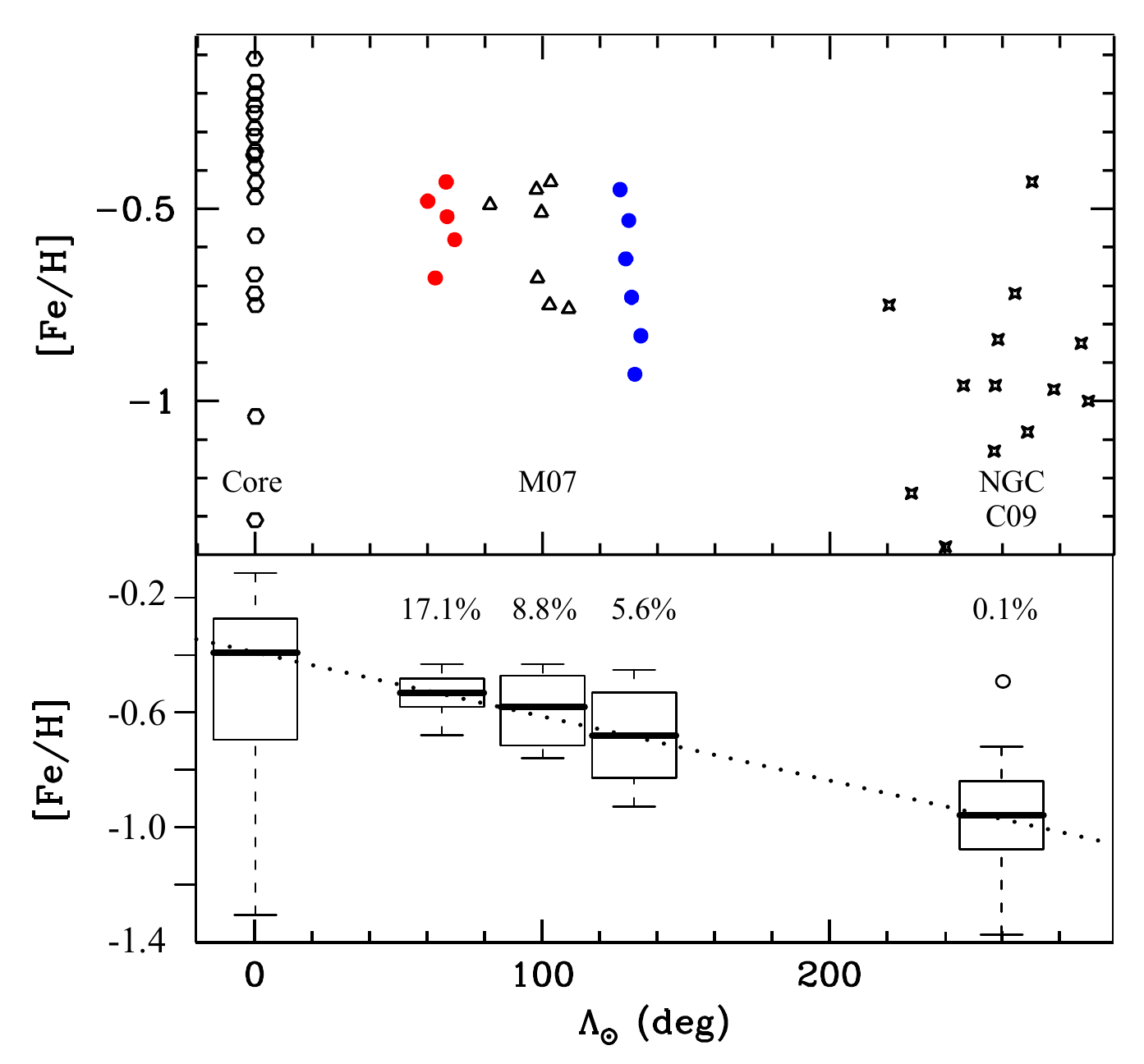}
\caption{[Fe/H] as a function of angular distance from the main body of Sgr along the trailing arm. The upper panel shows the individual points. In the lower panel the distribution of [Fe/H] in each sample is displayed as a box plot in which the solid line represents the median, the boxed region spans the first to third quartiles (i.e. the interquartile range), and the bars represent $\pm$1.5 $\times$ the interquartile range. The core sample is taken from \citet{Monaco05}. The \citet{Monaco07} study, centred at $\Lambda_{\odot}$=100$^{\circ}$, is derived from a 27.5$^{\circ}$ wide region of the trailing arm. The `North Galactic Cap' sample is that from \citet{Chou09}. The percentage above the various groups along the trailing arm records the probability that each sample is drawn from the Sgr core sample (see text for details). The dotted line shows the result of a least squares linear fit to the core, $\Lambda_{\odot}$=66$^{\circ}$ and $\Lambda_{\odot}$=132$^{\circ}$ samples. }\label{fig:FeH}
\end{center}
\end{figure}

In Figure \ref{fig:FeH} we present our [Fe/H] determinations for the two sample regions in the upper panel of Figure \ref{fig:FeH}. In the lower panel of Figure \ref{fig:FeH} we graphically represent the median [Fe/H] and its interquartile range as a function of angular distance ($\Lambda_{\odot}$ as defined in \citet{Law05}) from the main body of Sgr. A two-sided K-S test is used to determine the probability that the stream samples were drawn from the core sample \citep{Monaco05}. The percentages above each sample in the lower panel of Figure \ref{fig:FeH} show the probability that this is the case. Low values of this probability indicate that the assumption that the stream samples are similar to that of the core is a poor one. More distant material is seen to be progressively less like the core sample. A gradient of [Fe/H]=$-(2.4\pm0.3)\times10^{-3}$ dex/degree is determined from a least-squares fit to the core, $\Lambda_{\odot}$=66$^{\circ}$ and $\Lambda_{\odot}$=132$^{\circ}$ samples. At a mean distance of 22 kpc this projects to $-(9.4\pm1.1)\times10^{-4}$ dex / kpc.

The target stars have been taken from the study of \citet{Majewski03} and are selected therein on the basis of their 2MASS colours as M-giant stars. The judicious selection of \citeauthor{Majewski03} isolates the upper red giant branch (RGB) of Sgr with low contamination from the Milky Way (MW) field. However, it also imposes a bias towards metal-rich stars as detailed in \citet{Majewski03}. Metallicities of [Fe/H]$ < -1$ dex are essentially excluded due to this colour selection. To minimise the effects of this imposed metallicity bias on our findings, the above figures compare our results with literature data that impose identical colour selection of the M-giants. Utilisation of M-giants also imposes an age range to the sample of stars we study. An M-giant of [Fe/H] = $-0.4$ dex (typical of the Sgr core; \citeauthor{Monaco05} \citeyear{Monaco05}) possesses an age of 2--2.5 Gyr. At lower metallicities an older age is required to reach the same $J$$-$$K$ color (for example a 1M$_{\odot}$ [Fe/H] = -0.4 star is 0.1 dex younger than a [Fe/H] = -0.7 star at $J$$-$$K$=1.0; \citeauthor{PadovaIso} \citeyear{PadovaIso}).

Our results may be compared to the metallicity gradient observed in the more extensively studied leading arm material. As noted above, the leading arm is more extensively phased mixed. That is to say the material lost in successive orbits is not as spatially differentiated as in the trailing arm. This effect would be expected to reduce the apparent metallicity gradient along the leading arm compared to the trailing arm. In their study of the leading arm M giants, \citet{Chou07}, report the metallicity distribution function (MDF) in two regions; one centred at $\Lambda_{\odot} \sim 230^{\circ}$ of around 100$^{\circ}$ in extent, and another region at $\Lambda_{\odot} \sim 30^{\circ}$ (proposed to be old leading arm debris displaced $\sim 390^{\circ}$ from the main body). The mean metallicities are found to be $-0.7$ dex and $-1.1$ dex respectively. Taken together with the mean metallicity of the core, this equates to a metallicity gradient of $-2.2\times10^{-3}$ dex/degree. This is compatible with the present results for the trailing arm.

The [Fe/H] gradient we derive here is also compatible with the mean metallicity determined, again from M-giants, in the sample of \citet{Monaco07} (marked M07 in Figure \ref{fig:FeH}). Further, it is noteworthy that our observed [Fe/H] gradient continues to the \citet{Chou07} 'North Galactic Cap positive velocity' group which is ascribed by \citeauthor{Chou07} to an old wrap of the trailing arm. The North Galactic Cap sample is not used in our determination of the metallicity gradient of the trailing arm since there is possible confusion with other kinematically distinct substructures in the direction of the North Galactic Cap sample. The North Galactic Cap sample occupies an area of the sky in which there is both leading and trailing arm material as well as material from the Virgo Stellar Stream \citep{Duffau06, Vivas08, Kellerb09, Prior09}. The dynamical models of \citet{Law05} predict that while the leading arm material at the position of the North Galactic Cap sample will possess negative velocities, the trailing material will possess velocities of $100<V_{GSR}<200$ kms$^{-1}$. The Virgo Stellar Stream material possesses a colder velocity profile centred at $V_{GSR}\sim 100$ kms$^{-1}$. \citet{Prior09} shows that considerable overlap in radial velocity exists between the two systems and consequently it is not clear how much of the North Galactic Cap sample is due to the Virgo Stellar Stream. \citet{Prior09b} finds that the metallicity of the RR Lyrae members of the Virgo Stellar Stream are uniformly metal-poor at [Fe/H]$\sim-1.7$. At such low metallicity few objects would be contributed to the 2MASS colour selection discussed above. However, the RR Lyraes of the Virgo Stellar Stream represent the old stellar population and nothing is known of the metallicity distribution of other populations within this halo substructure.

\subsection{Implications of the metallicity gradient in Sgr debris}
We now discuss the implications of our findings of a metallicity gradient in the stars along the trailing arm of Sgr in the context of models of the chemodynamic evolution of Sgr. Dynamical models of the tidal disruption of Sgr have been presented by numerous authors. \citet{Ibata01, Helmi04, Martinez-Delgado04, Law05} present models that seek to account for the distance to, and radial velocity of, the M-giants of \citet{Majewski03}. \citet{Fellhauer06} and \citet{Martinez-Delgado07} include further observational constraints from the SDSS regarding the distance to the Sgr leading arm. A common feature of these studies is that while they manage to qualitatively match the features of the stream, they do not provide a consistent model for the shape of the Milky Way's dark matter halo. In particular, the leading material is best matched by an oblate halo, whereas the trailing material is best matched by a prolate figure \citep{Prior09b,Newberg07,Yanny09,Law05}. \citet{Law09} demonstrates that by adopting a triaxial halo model, rather than the axisymetric models assumed in the studies above, a concordant solution is achievable. However, as \citet{Law09} point out the solution is somewhat unsatisfactory since such a configuration is expected to be dynamically unstable. 

While the above uncertainties remain in the modelling of the orbit of Sgr, the pertinent features for the present study are that stars are released preferentially during perigalactic passage with an orbital period of $\sim0.85$ Gyr \citep{Law05}. Consequently, our sample at $\Lambda_{\odot}$=66$^{\circ}$ was lost from Sgr on the present perigalactic passage approximately 0.5 Gyr ago, and the $\Lambda_{\odot}$=132$^{\circ}$ sample was lost $\sim1.3$ Gyr ago on the previous passage of Sgr \citep[see figure 1 of][]{Law05}.

Studies of the star formation history of the main body of Sgr \citep{Layden00,Siegel07} have revealed a complex and protracted star formation history with three major phases. These studies find an old (11 Gyr) metal-poor population of [Fe/H] $\sim -1.3$ dex, a dominant intermediate age population (6--4.5 Gyr, with possible bursts) with [Fe/H] $\sim -0.6$ dex and a young (2--3 Gyr) population of [Fe/H] $\sim -0.4$ to $-0.1 $ dex. If the age-metallicity relation (AMR) of Sgr was spatially uniform then we would expect our samples of M-giants to possess the metallicity appropriate for their relatively young age ($\sim$2--3 Gyr), namely $-0.4 <$ [Fe/H] $< -0.1$ dex (as dominates the core sample). Furthermore we would expect a negligible metallicity gradient in the debris stream. 

Our observations imply that the progenitor of the present-day Sgr did not possess a spatially uniform AMR. Seen another way, the time between perigalacticons does not allow sufficient time for chemically homogeneous in-situ elevation of the mean [Fe/H] to the levels we observe between successive orbits. Rather, as recognised by \citet{Chou07}, the abundance gradient observed must arise due to the stripping of the outer regions of the Sgr progenitor over which a metallicity gradient (and/or concomitant age gradient) was present.

Indeed such population gradients are typically observed in dwarf galaxies. As reviewed by \citet{Stinson09}, dwarf irregular galaxies universally show extended envelopes dominated by old RGB stars. Such haloes of old--intermediate age stars are seen across a range of galaxy luminosity and under a range of tidal conditions. Stellar population gradients are notably less distinct amongst the local dSph galaxies. However, gradients in the morphology of the horizontal branch are not uncommon and are indicative of metallicity and/or age gradients \citep{Harbeck01}.

It is plausible therefore, that the original Sgr progenitor possessed an extended and correspondingly metallicity segregated, halo consistent with our observations. As Sgr experienced strong tidal interaction with the Milky Way, the tidal radius of Sgr decreased with each successive orbit, leading to the loss of stars from outer regions of the progenitor. In this picture, it is the intermediate age M-stars (having arisen in perhaps the most recent star formation episode) of this outer Sgr halo population that we see in the tidally disrupted debris tails. The radial metallicity gradient imprinted by previous chemodynamical evolution gives rise to the metallicity gradient we observe here. Since the stars of the present study are amongst the most recent to form in Sgr, the fact that they exhibit a [Fe/H] gradient requires that the abundance gradient is long-lived, having survived for approximately a Hubble time. Moreover, our findings imply that radial mixing of enriched material must be of little significance.

\section{Abundance Ratio Gradients}

\begin{figure}
\begin{center}
\includegraphics[width=84mm]{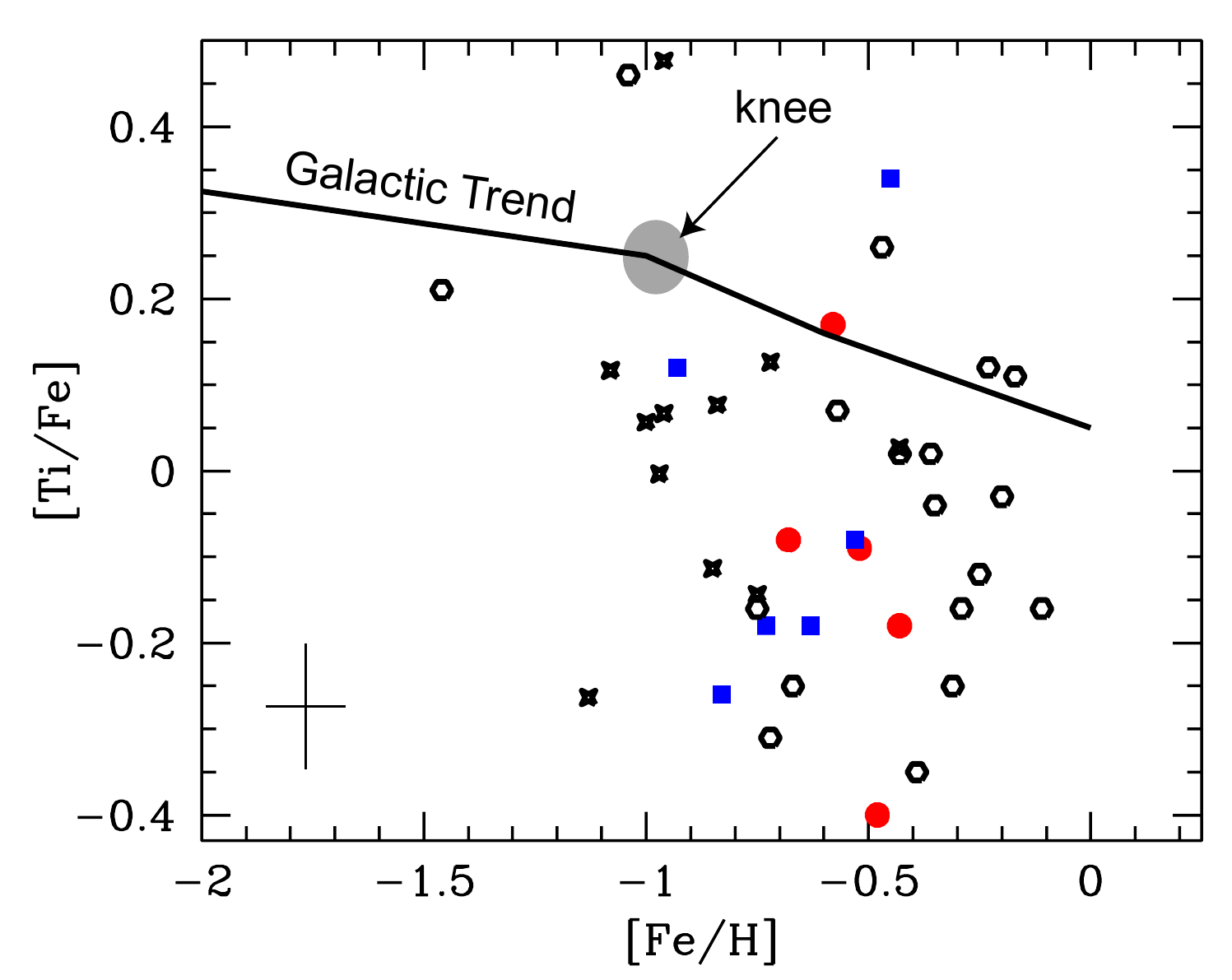}
\caption{[Ti/Fe] vs. [Fe/H]. Sgr core ({\it{open circles}}: \citet{Monaco05}), trailing arm sample at $\Lambda_{\odot}$=66$^{\circ}$ ({\it{red circles}}), trailing arm sample at $\Lambda_{\odot}$=132$^{\circ}$ ({\it{blue squares}}), and North Galactic Cap sample ({\it{crosses}}: \citet{Chou09}) are shown. For reference the Galactic locus is shown ({\it{solid line}}) as described in \citet{Venn04}. A representative error bar is shown.}\label{fig:TiFe}
\end{center}
\end{figure}

\begin{figure}
\begin{center}
\includegraphics[width=84mm]{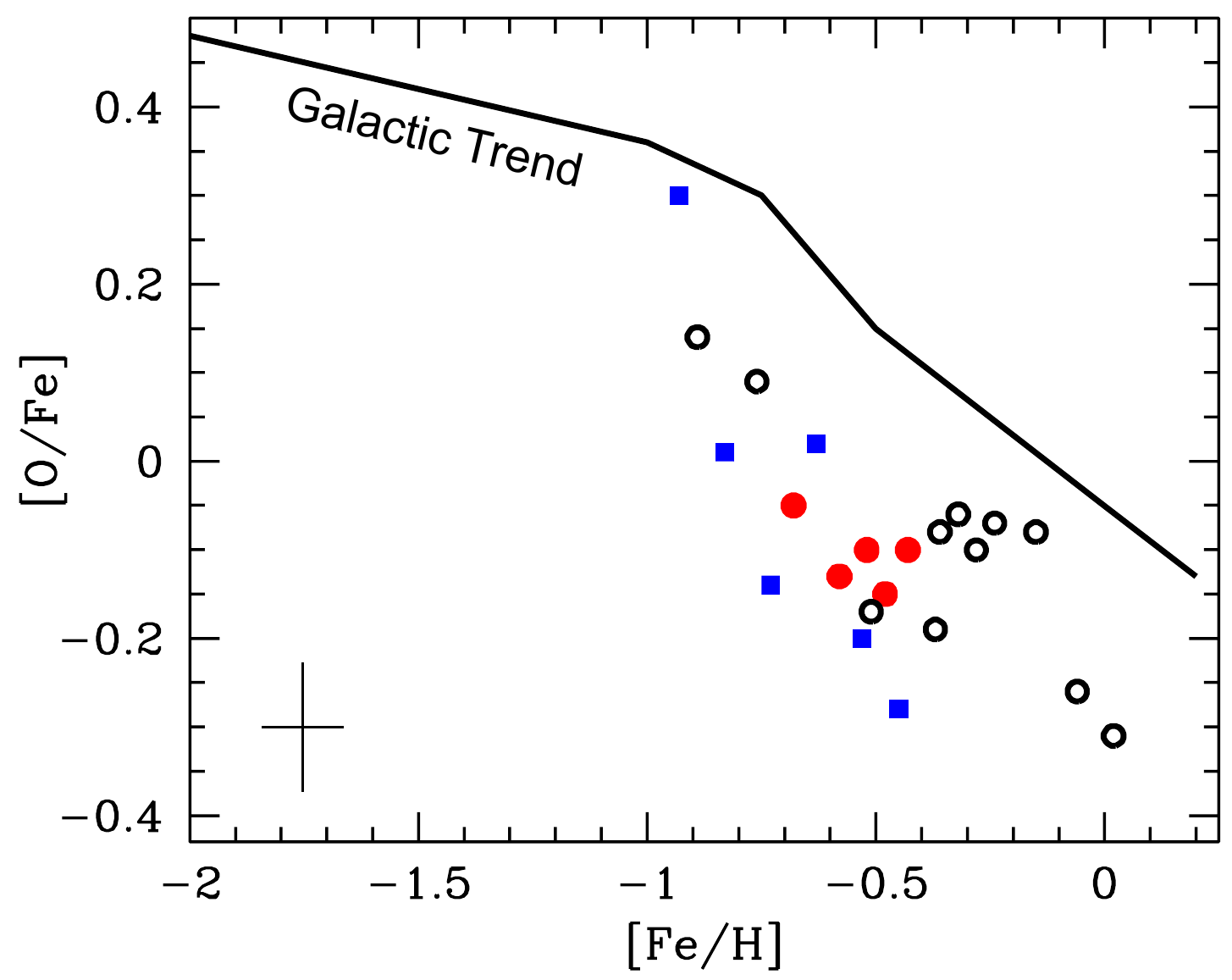}
\caption{[O/Fe] vs. [Fe/H]. Sgr core ({\it{open circles}}: \citet{Sbordone07}), trailing arm sample at $\Lambda_{\odot}$=66$^{\circ}$ ({\it{red circles}}), and trailing arm sample at $\Lambda_{\odot}$=132$^{\circ}$ ({\it{blue squares}}) are shown. For reference the Galactic locus is shown ({\it{solid line}}) as described in \citet{Venn04}. A representative error bar is shown.}\label{fig:OFe}
\end{center}
\end{figure}

In Figures \ref{fig:TiFe} and \ref{fig:OFe} we show our findings for [Ti/Fe] and [O/Fe] as a function of [Fe/H]. Both figures show a characteristic underabundance of O and Ti at given [Fe/H] relative to the stars of the Galactic disk and halo. We recall that the M giants of the present study represent stars of $\sim$ 2--3 Gyr; Galactic disk stars of this age reside on the Galactic locus (the solid line in Figs.\ \ref{fig:TiFe} \& \ref{fig:OFe}) at approximately Solar metallicity. 

\begin{figure}
\begin{center}
\includegraphics[width=84mm]{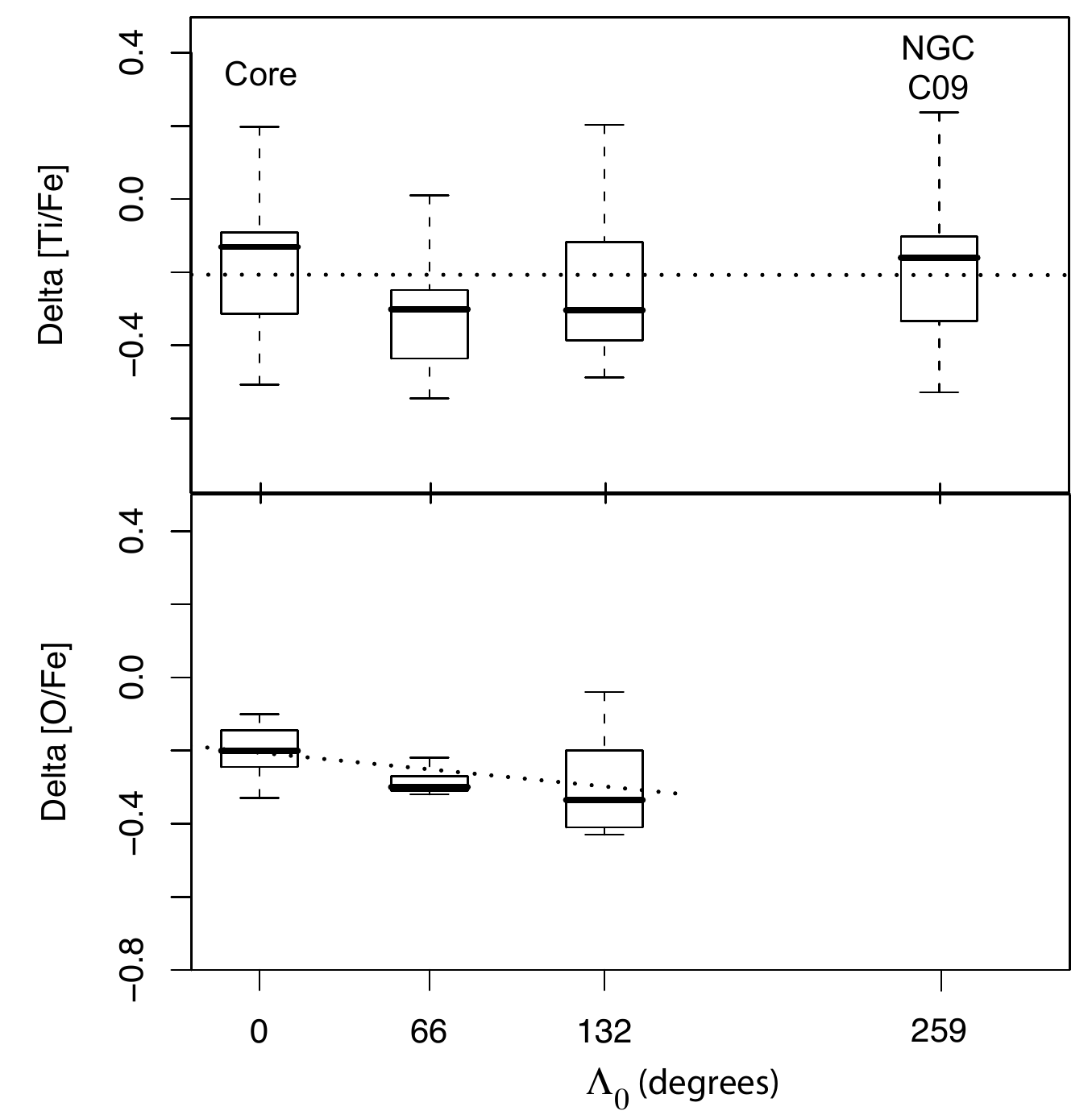}
\caption{{\bf{Top}}: The offset in [Ti/Fe] at a given [Fe/H] relative to the Galactic halo+disk locus \citep{Venn04} as a function of angular distance from the main body of Sgr. The dotted line shows the least squares linear fit to the samples. Other details of the figure are as described in Fig.\ \ref{fig:FeH}. No spatial gradient in [Ti/Fe] is apparent, however, the spread of [Ti/Fe] at given [Fe/H] is large. {\bf{Bottom}}: Same as above for [O/Fe]. }\label{fig:TiFe_OFe}
\end{center}
\end{figure}

Figure \ref{fig:TiFe_OFe} shows the interquartile ranges of the offset in [O,Ti/Fe] at given [Fe/H] between our observed sample and that defined in the Galactic disk and halo\footnote{ in the sense [$X$/Fe]$_{Sgr}$ - [$X$/Fe]$_{Galactic}$, where $X$ is the abundance of O or Ti.} as a function of angular distance from Sgr. The probability that each sample is drawn from the core sample is determined by a two-sided K-S test. In the case of both [O/Fe] and [Ti/Fe] there is no significant gradient along the trailing arm. However, a significant spread in [Ti/Fe] is apparent in the four samples in Figure \ref{fig:TiFe}. The spread is much larger than the abundance ratio errors (the associated abundance ratio errors of the studies of \citet{Monaco05} and \citet{Chou09} are similar in magnitude to those of the present study). Such a spread is not apparent in [O/Fe], shown in Figure \ref{fig:OFe}. The origin of this spread in [Ti/Fe] is not understood. Our study does not however, enable us to define the [Fe/H] of the [O/Fe] or [Ti/Fe]  `knee'. This is due to the inherent metal-rich bias in our use of M-giants (as discussed above) and small sample size. Consequently, we do not constrain the expected metal-poor ([Fe/H]$< -1$ dex) [O/Fe] plateau. Chemical evolution models of \citet{Lanfranchi04} show that the progressively lower SF efficiency should lead to a [$\alpha$/Fe] plateau at progressively lower [$\alpha$/Fe] at increasing $\Lambda_{\odot}$. Abundances of radial-velocity selected K-giants (representative of the metal-poor population) in the trailing arm would enable clarification of the location of the `knee' in the [$\alpha$/Fe] relation, or otherwise, of a chemical gradient in the trailing arm material. 

\subsection{Implications of $\alpha$-element abundance}
Our observations of the $\alpha$-element O and the $\alpha$-like element Ti, show that the [$\alpha$/Fe] ratio of Sgr material is lower at a given [Fe/H] compared with that of the MW (at least for [Fe/H] $> -1$ dex material probed to date). As described in a large literature of chemical evolution models \citep[see for example,][]{Gilmore91, Lanfranchi04, Lanfranchi06, Matteucci08} and motivated by numerous observational studies \citep[for example,][]{Smith02,Venn04,Cohen09,Chou09,Hidalgo09,Lee09} the generally lower [$\alpha$/Fe] is a consequence of slower chemical enrichment in lower mass systems. The slower pace of chemical enrichment (due to lower star formation efficiency and/or stronger galactic winds) allows Fe-rich SNIa products to be incorporated at lower metallicities relative to the MW halo.

The [Fe/H] of the [$\alpha$/Fe] `knee' feature, seen in the MW halo at a [Fe/H] $\sim -1.0$ dex, provides a snapshot of the metallicity at $\sim1$ Gyr, after which time SNIa begin to appreciably raise the Fe content and reduce [$\alpha$/Fe]. As suggested by \citet{Monaco05}, and demonstrated by \citet{Chou09} in the Sgr leading arm, for [$\alpha$/Fe] $\leq -1$ dex there is a general overlap in the [$\alpha$/Fe] ratio of MW and Sgr stars. This indicates a common dominance of SNII enrichment at these metallicities. For [Fe/H] $\geq -1$ dex however, the [$\alpha$/Fe] ratio of Sgr material is significantly below that the of the MW. This possibly reflects a lower early star formation rate (SFR) in Sgr compared to the MW, or as proposed by \citet{Lanfranchi06}, that Sgr exhibited a moderate star formation (SF) efficiency at first, but one that was quenched by galactic winds from ensuing SNII.

The implications of the evolution of [$\alpha$/Fe] in Sgr can be contrasted with those in the LMC. In the LMC the [Ti/Fe] and [O/Fe] ratio remains lower that the MW at all metallicities probed to date ([Fe/H]$\lesssim$-1.3 \citep{Hill00, Smith02}. This is ascribed to a SF efficiency lower than that of Sgr and/or more efficient galactic winds (discussed in the context of [O/Fe] by \citet{Gilmore91} \& \citet{Smith02}) and the presence of fewer contributing high-mass SNII (from [Ti/Fe], \citet{Pompeia08}).

\citet{Monaco05} therefore point out that the early chemical evolution of Sgr was more akin to that of the MW than that of the LMC and the local dSphs. They propose that the progenitor of Sgr was most likely a gas-rich, star forming galaxy that was substantially more massive that the extant remnant. The \citet{Lanfranchi06} results amplify this argument. The study of \citeauthor{Lanfranchi06} defines a sequence of relative SFR in early galaxy evolution that ranges from the dSphs which exhibit the lowest SFR, to the LMC with a moderate SFR, and finally Sgr with the highest SFR evident amongst the MW satellites. Thus the advanced chemical evolution of Sgr material argues strongly that the progenitor was more massive that the LMC. However, at present the dynamical models of \citet{Law05} assume a progenitor mass of 2--5$\times$10$^{8}$M$_{\odot}$ compared to the LMC mass of 1--2$\times$10$^{9}$M$_{\odot}$. An increase in the progenitor mass of Sgr would imply a more extended history of mass-loss to the MW and an enhanced impact on the chemistry of the Halo. We also note that a large mass progenitor would be in line with the observed extended SFH of Sgr that indicates that Sgr was actively forming stars until 1--2 Gyr ago \citep[][indeed if it had not, there would be no M giants with which to trace the tidal streams]{Siegel07} although it  has been tidally interacting with the MW for $\sim 3$ Gyr \citep[see for example][]{Law05}. The ability to retain the gas required for such star formation in the face of Galactic tidal interaction would argue for a large progenitor mass.

\subsection{Sgr and the formation of the Galactic Halo}
An important role is expected for accreted satellites in the formation of the Galactic Halo within Lambda-CDM cosmology. Here hierarchical mergers lead to massive galaxies surrounded by a flotilla of lower mass satellite halos (at least as traced by dark matter-only simulations). Hence, the stellar content of the Halo must presumably contain some contribution from the disruption of such satellites. A number of observational challenges to such a scenario have arisen.

Firstly, the number of lower mass halos observed about the MW was found to be low compared to predictions of cosmological simulations (the so-called `missing satellite problem' \citeauthor{Klypin99} \citeyear{Klypin99}). However, recent searches have revealed increased numbers of extremely faint dSphs \citep{Walsh09,fieldofstreams}. When considered together with survey selection effects, the shortfall of local dSphs is substantially ameliorated \citep{Koposov09}.

Secondly, amongst metal-poor stars ([Fe/H $< -2.5$ dex) the MDF and chemistry of the Galactic halo were seen to be very different to those exhibited by dSphs \citep{Geisler07}. For example, the metal-poor tail of the MDF in dSphs was seen to be deficient compared to that of the Halo \citep{Helmi06}. However, subsequent studies of \citet{Starkenburg10} do not uphold this deficiency. \citet{Schorck09}, \citet{Cohen09} and \citet{Norris08} conclude that the MDF of dSphs and the Halo are in good agreement. Furthermore, evidence now shows that the extremely metal-poor stars in dSphs show similar [$\alpha$/Fe] to the Halo \citep{Frebel10a, Norris10}. It is therefore possible that present-day dSphs could have contributed to the metal-poor Halo.

However, divergent chemistry remains at higher metallicity. For [Fe/H]$>-1$, the [$\alpha$/Fe] of the dSph sample is uniformly, and significantly, lower than that of the Halo. Only Sgr, thought to be (in the form of its progenitor) the most massive of the local dwarf galaxies, has some stars with [$\alpha$/Fe] similar to the Halo at intermediate metallicities. Signature of slow chemical enrichment, the [$\alpha$/Fe] of an average low-mass dSph implies that such systems can not be responsible for a substantial fraction of the metal-poor halo. 

In order to explain the Halo chemistry, it is necessary to capture [Fe/H] $\gtrsim -2$ and $\alpha$-enhanced material. One mechanism to introduce such material is from a number of massive satellites that experienced rapid star formation and were then accreted rapidly, before the onset of SNIa contributions \citep{Robertson05, Font06, DeLucia08}. \citet{Lagos09} make a clear distinction between `building block' satellites (i.e.\ those that are accreted) and surviving satellites. From cosmological simulations, they find that 'building blocks' collapse and form stars earlier than surviving satellites that instead form stars in a quiescent manner. \citet{Cooper09} finds that the merger of less than five massive ($10^8$ M$_{\odot}$) `building blocks' at early times ($1<z<7$) can account for the Galactic halo.

Our findings show that the chemistry of the stars contributed by Sgr to the Galactic halo in the trailing arm debris are significantly different from that of the halo at similar [Fe/H]. Therefore Sgr is not a `building block'. The debris are increasingly metal-poor further from the mainbody of Sgr and so, in terms of the [Fe/H] the material is increasingly `halo-like'. However, we see no signs of convergence between the [$\alpha$/Fe] ratio in stars of the trailing arm and stars of similar [Fe/H] from the Galactic Halo. Consequently, our findings show that the fraction of the intermediate -- metal-rich halo population that can have been contributed by the dissolution of Sgr-like objects over the last several Gyr can not be substantial. 

\section{Summary}
In this study we have presented the  abundance analysis of eleven M-giant stars in two regions along the trailing arm of the Sgr dwarf. The two populations, together with existing data from the literature, enable us to explore the metallicity and chemistry of stars deposited into the Milky Way halo from Sgr over the past $\sim1.3$ Gyr. Within our limited sample of $\sim$2--3 Gyr old stars, we find a significant gradient in metallicity along the debris stream. The metallicity gradient is $-(2.4 \pm 0.3) \times 10^{-3}$ dex / degree ($-(9.4\pm1.1)\times10^{-4}$ dex / kpc) decreasing away from the main body of Sgr. The [$\alpha$/Fe] ratio at a given metallicity is seen to be approximately constant along the extent of the trailing arm. The change in metallicity along the stream can be understood to be due to the tidal disruption of a Sgr progenitor that possessed a radial gradient in [Fe/H]. Such a gradient could have arisen due to a radial decrease in the star formation efficiency and/or a radial increase in the efficiency of galactic wind-driven metal loss. The physical mechanism that produced a radial [Fe/H] gradient that persisted over a Hubble time did not, however, produce a detectable gradient in [$\alpha$/Fe].

There is no significant [$\alpha$/Fe] gradient along the trailing arm from our sample of M-giants. The [$\alpha$/Fe] ratios of our targets is distinct from the halo.  It can therefore be concluded that while some portion of the metal-poor halo might be contributed via the accretion of Sgr-like objects, the metal-rich component of the Halo can not feature a substantial contribution from such objects over the last several Gyr.

\acknowledgments

We thank Jorge Mel{\' e}ndez for providing the line list. This research has been supported in part by the Australian Research Council through Discovery Project Grants DP0343962 and DP0878137. Based on observations obtained at the Gemini Observatory for Program GS-2008B-Q-33, which is operated by the Association of Universities for Research in Astronomy, Inc., under a cooperative agreement with the NSF on behalf of the Gemini partnership: the National Science Foundation (United States), the Science and Technology Facilities Council (United Kingdom), the National Research Council (Canada), CONICYT (Chile), the Australian Research Council (Australia), Ministerio da Ciencia e Tecnologia (Brazil)  and Ministerio de Ciencia, Tecnologia e Innovacion Productiva  (Argentina).





{\it Facilities:} \facility{Gemini:South (Phoenix)}


\clearpage



\clearpage

\clearpage

\clearpage

\clearpage

\clearpage

\clearpage

\clearpage

\begin{deluxetable}{l c c c c c c c c c c }
\tabletypesize{\scriptsize}
\rotate
\tablecaption{Log of {\it Phoenix} observations. The first five stars correspond to the `66 degree' group, and the remaining six correspond to the `132 degree' group discussed in the text.\label{table:obslog}}
\tablewidth{0pt}
\tablehead{
\colhead{Star} & \colhead{RA (J2000)} & \colhead{Dec (J2000)} & \colhead{$\Lambda_{\odot}$ (degrees)} & \colhead{$J$} &
\colhead{$K$} & \colhead{E($B$$-$$V$)} & \colhead{($J$$-$$K$)$_{0}$\tablenotemark{a}} &
\colhead{UT Date} & \colhead{Exposure (seconds)} &
\colhead{S/N}
}
\startdata
2329301m245810 & 23:29:30.1 & -24:58:10 & 60.00 & 11.590 & 10.499 & 0.02 & 1.145 & 2008-08-21 & 6 $\times$ 562 & 94\\
2345417m264456 & 23:45:41.7 & -26:44:56 & 62.63 & 11.413 & 10.393 & 0.02 & 1.072 & 2008-08-21 & 6 $\times$ 562 &98\\
2350361m200216 & 23:50:36.1 & -20:02:16 & 66.41 & 11.664 & 10.576 & 0.02 & 1.142 & 2008-09-17 & 6 $\times$ 562 & 68\\
2353194m205041 & 23:53:19.4 & -20:50:41 & 66.72 & 12.571 & 11.621 & 0.02 & 1.000 & 2008-09-17 & 12 $\times$ 562 & 65\\
0003528m194047 & 00:03:52.8 & -19:40:47 & 69.35 & 12.024 & 11.007 & 0.03 & 1.069 & 2008-09-19 & 6 $\times$ 562 & 52\\
0334210p051809  & 03:34:21.0 & +05:18:09 & 126.93 & 12.505 & 11.345 & 0.26 & 1.355 & 2008-09-19 & 12 $\times$ 562 & 110\\
0340164p090338 & 03:40:16.4  & +09:03:38 & 130.03 & 12.971 & 11.757 & 0.35 & 1.462 & 2009-01-11 & 12 $\times$ 562 & 30\\
0342225p054745 & 03:42:22.5 & +05:47:45 & 128.93 & 11.976 & 10.819 & 0.22 & 1.328 & 2008-08-22 & 6 $\times$ 562 &78\\
0348437p065236 & 03:48:43.7 & +06:52:36 & 130.97 & 11.976 & 10.836 & 0.20 & 1.299 & 2009-02-02 & 6 $\times$ 562 &97 \\
0357262p053258 & 03:57:26.3 & +05:32:48 & 132.12 & 11.375 & 10.130 & 0.29 & 1.459 & 2008-12-19 & 6 $\times$ 562 &79 \\
0408285p044043 & 04:08:28.5 & +04:40:04 & 134.19 & 12.681 & 11.457 & 0.40 & 1.501 & 2008-09-19 & 6 $\times$ 562 &65 \\
\enddata
\tablenotetext{a}{Dereddened color in the Johnson-Glass system.}
\end{deluxetable}


\clearpage

\begin{table}
\begin{center}
\caption{Derived stellar parameters for target stars.\label{table:starpars}}
\begin{tabular}{l c c c }
\tableline\tableline
Star & $T_{{\rm{eff}}} (K)$ & log $g$ & $\xi_{t}$ (km\ s$^{-1}$)\\
\tableline
2329301m245810 & 3531 & 0.03 & 2.08\\
2345417m264456 & 3665 & 0.22 & 2.01\\
2350361m200216 & 3536 & 0.09 & 2.08\\
2353194m205041 & 3807 & 1.16 & 1.92\\
0003528m194047 & 3670 & 0.22 & 2.00\\
0334210p051809  & 3654 & 0.22 & 2.00\\
0340164p090338 & 3646 & 0.16 & 2.01\\
0342225p054745 & 3616 & 0.11 & 2.03\\
0348437p065236 & 3627 & 0.13 & 2.02\\
0357262p053258 & 3526 & 0.02 & 2.08\\
0408285p044043 & 3683 & 0.23 & 1.99\\
\tableline
\end{tabular}
\end{center}
\end{table}

\clearpage

\begin{table}
\begin{center}
\caption{Derived stellar abundances for target stars.\label{table:abundances}}
\begin{tabular}{l c c c c}
\tableline\tableline
Star &$A$(C) & $A$(O) & $A$(Ti) & $A$(Fe)\\
\tableline
2329301m245810 & 7.27 & 8.03 & 4.02 & 6.97\\
2345417m264456 & 7.19 & 7.93 & 4.14 & 6.77\\
2350361m200216 & 7.34 & 8.13 & 4.29 & 7.02\\
2353194m205041 & 7.31 & 8.03 & 4.30 & 6.93\\
0003528m194047 & 7.01 & 7.95 & 4.49 & 6.87\\
0334210p051809  & 7.21 & 7.93 & 4.79& 7.00\\
0340164p090338 & 7.21 & 7.93 & 4.29 & 6.92\\
0342225p054745 & 6.91 & 8.05 & 4.09 & 6.82\\
0348437p065236 & 6.96 & 7.79 & 3.99 & 6.72\\
0357262p053258 & 7.24 & 8.03 & 4.09 & 6.52\\
0408285p044043 & 7.09 & 7.84 & 3.81 & 6.62\\
\tableline
\end{tabular}
\tablecomments{$A$(X) = log[$n$(X)/$n$(H)] + 12. For reference the \citet{Asplund06} Solar values of the above elemental abundances are: $A$(C)=8.39, $A$(O)=8.66, $A$(Ti)=4.90, and $A$(Fe)=7.45.}
\end{center}
\end{table}

\clearpage

\begin{table}
\begin{center}
\caption{Stellar abundance dependencies on model parameters for 0348437p065236}
\label{table:abundanceUncertainties}
\begin{tabular}{l c c c c}
\tableline\tableline
Species &$T_{\rm{eff}}$ $\pm$ 75K & log $g$ $\pm$ 0.4dex & $\xi_{t}$ $\pm$ 0.4km\ s$^{-1}$ & Total\tablenotemark{a}\\
\tableline
$\Delta A$(C) $\ldots$ & 0.08 & 0.09 & 0.06 & 0.13\\
$\Delta A$(O) $\ldots$ & 0.12 & 0.10 & 0.06 & 0.15\\
$\Delta A$(Ti) $\ldots$ & 0.08 & 0.08 & 0.09 & 0.15\\
$\Delta A$(Fe) $\ldots$ & -0.08 & 0.04 & -0.04 & 0.10\\
\tableline
\end{tabular}
\tablenotetext{a}{The total value is the sum in quadrature of the individual abundance dependencies.}
\end{center}
\end{table}


\end{document}